\documentclass[prl,floatfix,superscriptaddress,twocolumn,showpacs]{revtex4-1}
\usepackage{graphicx}
\usepackage{amsmath,amssymb,amsfonts}
\usepackage{mathtools}
\usepackage{bm}
\usepackage{color}
\usepackage[colorinlistoftodos]{todonotes}

\begin{document}
\title{Aging in thermal active glasses}
\author{Giulia Janzen}
\author{Liesbeth M.~C.~Janssen}
\email{l.m.c.janssen@tue.nl}
\affiliation{Department of Applied Physics, Eindhoven University of Technology, P.O.~Box 513, 5600 MB Eindhoven, The Netherlands}

\date\today

\begin{abstract}
It is well established that glassy materials can undergo aging, i.e., their properties gradually change over time. There is rapidly growing evidence that dense active and living systems also exhibit many features of glassy behavior, but it is still largely unknown how physical aging is manifested in such active glassy materials. Our goal is to explore whether active and passive thermal glasses age in fundamentally different ways. To address this, we numerically study the aging dynamics following a quench from high to low temperature for two-dimensional passive and active Brownian model glass-formers. 
We find that aging in active thermal glasses is governed by a time-dependent competition between thermal and active effects, with an effective temperature that explicitly evolves with the age of the material. Moreover, unlike passive aging phenomenology, we find that the degree of dynamic heterogeneity in active aging systems is relatively small and remarkably constant with age. We conclude that the often-invoked mapping between an active system and a passive one with a higher effective temperature rigorously breaks down upon aging, and that the aging dynamics of thermal active glasses differs in several distinct ways from both the passive and athermal active case.
\end{abstract}

\maketitle
\textit{Introduction --} Glasses are disordered solids that exhibit extremely slow relaxation dynamics \cite{biroli2013perspective,debenedetti2001supercooled,langer2014theories}. An important hallmark of glasses is that they undergo physical aging, i.e., the behavior of the material depends explicitly on its age \cite{hodge1995physical,berthier2009statistical,lunkenheimer2005glassy,zhao2013using}. In supercooled liquids and glasses, aging dynamics is often studied by applying a temperature quench toward a lower temperature; after such a quench, the structural relaxation time tends to increase with the waiting time (or 'age' of the material) $t_w$ \cite{kob1997aging,foffi2004aging}. % in a power-law fashion. 
In general, this aging behavior can be regarded as an out-of-equilibrium phenomenon whereby the material seeks to recover equilibrium at the new temperature \cite{struik1977physical,hutchinson1995physical,barrat1999fluctuation,kob2000fluctuations}. More specifically, physical aging of thermal glasses is qualitatively understood as the gradual approach toward a lower-lying energy state on the energy landscape \cite{debenedetti2001supercooled}. 

The recent advent of active matter, i.e., systems whose constituent particles can move autonomously through the consumption of energy, is fueling renewed interest in the study of glassy dynamics \cite{janssen2019active,berthier2019glassy}. 
Both theory and simulations \cite{szamel2015glassy,szamel2016theory,feng2017mode,liluashvili2017mode,szamel2019mode,henkes2011active,ni2013pushing,berthier2013non,berthier2014nonequilibrium,szamel2015glassy,flenner2016nonequilibrium,berthier2017active}, as well as recent experiments of synthetic systems \cite{klongvessa2019active,klongvessa2019nonmonotonic} and living cells \cite{zhou2009universal,parry2014bacterial,garcia2015physics,bi2016motility,nishizawa2017universal,angelini2011glass}, have established that dense active matter can show remarkable similarities with conventional glassy phenomenology. For example, slow structural relaxation and ultimate kinetic arrest, varying degrees of fragility, Stokes-Einstein violation, and dynamic heterogeneity have all been observed in disordered active systems \cite{ediger2000spatially,berthier2011theoretical,binder2011glassy,berthier2005direct,richert2002heterogeneous,ediger2012perspective,janssen2019active}. 
A popular model for such active glassy dynamics is the active Brownian particle (ABP) model, which combines thermal diffusive motion with a constant self-propulsion speed for each particle \cite{romanczuk2012active,ramaswamy2017active,ramaswamy2010mechanics,lindner2008diffusion,lowen2020inertial,sprenger2020active}. 
The prevalent qualitative picture of ABP glassy dynamics  is that -- at least in steady-state conditions -- the presence of active forces generally facilitates cage-breaking events at high densities \cite{ni2013pushing}. This culminates in overall faster relaxation dynamics, and hence ABPs are often mapped onto an effective passive system with a higher effective temperature \cite{howse2007self,bechinger2016active}.

In contrast to many other hallmarks of glassy dynamics, the manifestation of physical aging in active glasses is still largely unexplored. \textit{A priori} it is not clear how an active glass will age, as the conventional picture of aging, i.e., the material's tendency to reach equilibrium via minimization of the energy, cannot hold in an active non-Hamiltonian system that is inherently far from equilibrium \cite{vicsek2012collective,marchetti2013hydrodynamics,speck2016collective,de2013non,lebon2008understanding}. Moreover, since aging, activity, and glass formation all constitute different types of non-equilibrium phenomena, one might expect a complex interplay of various dynamic processes within such a system. The limited number of active aging studies reported thus far have focused only on \textit{athermal} systems \cite{janssen2017aging,mandal2020multiple}, and hence it is still unclear to what extent  physical aging in thermal glasses becomes fundamentally different when introducing activity.

In this work, we explore how the role of temperature -- the key parameter in most conventional (passive) aging studies --, affects the aging dynamics of thermal active glasses following a temperature quench. We find that the active aging dynamics of ABPs is governed by a non-trivial and time-dependent competition between thermal and active effects, which generally precludes the mapping of the active system onto an effective passive one. In particular, activity leads to a relative speed-up in structural relaxation as time progresses, while the degree of dynamic heterogeneity remains remarkably constant with age. 
We attribute these findings to activity-enhanced cage breaking -- a reverberation of the well-known steady-state ABP dynamics \cite{ni2013pushing}. Overall, we conclude that the aging behavior of thermal active glasses is distinct from both the passive and athermal active case.

\textit{Simulation details --} We study a two-dimensional (2D) binary mixture of thermal ABPs. The overdamped equations of motion for each particle $i$ are given by
\begin{align}
   &\gamma \,\dot{\bm{r}}_i=\sum_{i\ne j=1}^{N} \bm{f}_{ij}+f \, \bm{n}_i+\sqrt{2D_T} \, \bm{\eta} \label{1}\\ 
 &\dot{\theta}_i=\sqrt{2D_{r}} \,\eta_{\theta} \label{2}
\end{align}
where $\bm{r}_i=(x_i,y_i)$ represents the particle's spatial coordinates, $\theta_i$ the rotational coordinate, and the dots denote the time derivative. The thermal noise is represented by independent white noise stochastic processes $\bm{\eta}=(\eta_x,\eta_y)$  with zero mean and variance $2k_BT/\gamma \delta(t-t^\prime)$, where $k_B$ is the Boltzmann constant, $T$ the temperature, and $\gamma$ a friction coefficient. In Eq.\ \ref{2}, $\eta_{\theta}$ is a Gaussian stochastic process with zero mean and variance $2 D_r \delta(t-t^\prime)$. The translational and rotational diffusion constants are denoted as $D_T$ and $D_{r}$, respectively, and we can further define the ABP persistence time as $\tau_r=D_r^{-1}$. Each ABP has a constant self-propulsion speed $f /  \gamma$ along a direction $\bm{n}_i=(\cos{\theta}_i,\sin{\theta}_i)$; if the magnitude of the active force, $f$, is zero, Eq.\ \ref{1} reduces to the equation of motion for a passive Brownian particle. Lastly, $\bm{f}_{ij}=-\nabla_i V(r_{ij})$ is the interaction force between particles $i$ and $j$, where $r_{ij}=\left|\textbf{r}_i-\textbf{r}_j \right|$. For $V$ we use a Lennard-Jones potential 
  \begin{equation*}
      V(r_{ij})=4\epsilon_{ij} \left[\left(\frac{\sigma_{ij}}{r_{ij}} \right)^{12} -\left(\frac{\sigma_{ij}}{r_{ij}} \right)^{6}    \right]
  \end{equation*}
with a cutoff distance $r_{ij} = 2.5 \sigma_{ij}$.
%and zero otherwise. Here $r_{ij}=\left|\textbf{r}_i-\textbf{r}_j \right|$, $\sigma_{ij}$. 
%and $\epsilon_{ij}$ depend on  the particle species (A or B). 
In order to prevent crystallization we use the parameters of the 2D binary Kob-Andersen mixture \cite{kob1995testing}: $A=65 \%$, $B=35 \%$,  $\epsilon_{AA}=1$, $\epsilon_{BB}=0.5 \epsilon_{AA} $, $\epsilon_{AB}=1.5 \epsilon_{AA}$, $\sigma_{AA}=1$, $\sigma_{BB}=0.88 \sigma_{AA}$ and $\sigma_{AB}=0.8 \sigma_{AA}$. 
We set the density to $\rho=1.2$, the number of particles to $N=10\,000$ and $D_T=\gamma=1$. Results are in reduced units, where $\sigma_{AA}$, $\epsilon_{AA}$, $\frac{\sigma_{AA}^2 \gamma}{\epsilon_{AA}}$, and $\frac{\epsilon_{AA}}{k_B}$ are  the units of length, energy, time, and temperature, respectively. The Brownian dynamics simulations were performed by integrating the equations of motion using the Euler-Maruyama method with a step size $\delta t=10^{-4}$. 

In order to reach the glass transition temperature $T_g$, we slowly decrease the temperature of the liquid starting from $T=1$. 
To reach the steady state, for each temperature $T$, we let the system equilibrate for a time larger than the structural relaxation time before collecting data. To study the aging behavior we prepare $100$ independent configurations and we let them equilibrate at the initial temperature $T_i=1$. After this equilibration process we apply a quench to the final temperature $T_q \lesssim T_g$ and then we follow the evolution in time at constant temperature ($T=T_q$). 
In the passive case we use quenching temperatures between $T_q=0.25$ and $T_q=0.4$, while for the active system ($f=0.5$, $D_r=0.1,\, 1,\,10$) we typically use $T_q=0.25$; we have chosen these values to admit comparisons between passive and active systems at both the same absolute temperature, and at the same quasi-effective temperature in steady state.
Finally, while here we consider the aging dynamics following a temperature quench, it may be expected that a qualitatively similar aging behavior applies when using an activity quench. We have verified this similarity for a few dynamical properties, but we will focus solely on the temperature-quench dynamics in the remainder of this work.

\begin{figure}
    \centering
    \includegraphics [width=9cm,height=3.3cm] {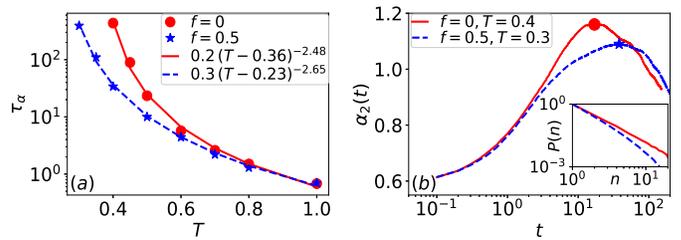} 
    \caption{(a): Temperature dependence of the structural relaxation time $\tau_{\alpha}$ for a passive and active system in steady state. In the passive case ($f=0$, red dots), we report $\tau_{\alpha}$ for $T=1,0.8,0.7,0.6,0.5,0.45,0.4$. In the active case ($f=0.5$ and $\tau_r=D_r=1$, blue stars), we plot $T=1,0.8,0.7,0.6,0.5,0.4,0.35,0.3$. Lines represent a fit to the power law $\tau_{\alpha}=\tau_0 \,  (T-T_c)^{-\Gamma}$. 
    (b): Non-Gaussian parameter $\alpha_2(t)$ as a function of time $t$ in steady state. The red line corresponds to the passive system at $T=0.4$ and the blue dashed line to the active system at $T=0.3$.
    The red dot and blue star indicate the respective peaks of $\alpha_2(t)$.  The inset shows the corresponding probability distributions $P(n)$ of the size $n$ of clusters of mobile particles around the peak of $\alpha_2(t)$.}
    \label{fig:equilibrium}
\end{figure}

\textit{Steady state dynamics --} Before discussing the aging dynamics, we first consider the steady-state dynamics as a benchmark.  %and identify the relevant parameter regimes. The structural relaxation time $\tau_{\alpha}$ can be extracted from the self intermediate scattering function via the definition $F(k,\tau_{\alpha})=e^{-1}$ \cite{berthier2011theoretical,janssen2018mode,flenner2015fundamental}.
In Fig.\ \ref{fig:equilibrium}(a) we show the temperature dependence of the $\alpha$ relaxation time for both a passive ($f=0$) and active ($f=0.5, \tau_r=D_r=1$) system. 
%here the relaxation is defined  via $F(k,\tau_{\alpha})=e^{-1}$ \cite{flenner2015fundamental}. 
We find that at any given temperature below $T\approx0.8$, the active system always tends to relax faster than the passive reference case, in agreement with literature \cite{mandal2016active,ni2013pushing,berthier2019glassy}. 
In particular, the temperature at which $\tau_{\alpha}$ reaches a value of $10^3$ is $T\approx0.4$ for the passive case, and $T\approx0.3$ for the active case. 
To estimate the glass transition temperature more precisely, we have also fitted both data sets with a power law $\tau_{\alpha}=\tau_0 \,  (T-T_c)^{-\Gamma}$, yielding $T_c=0.36$ and $T_c=0.23$ for the passive and active mixture, respectively. 
Note that, despite the high densities in both systems, the observed  differences in $T_c$ are fairly close to the expected effective temperature differences in the dilute limit \cite{bechinger2016active}: For a single ABP with activity parameters $f=0.5$ and $D_r=1$, a temperature of $T=0.25$ would correspond to a 'passive' effective temperature of $T_{eff}=0.375$.
%With these results we can also establish the appropriate temperature regimes for our remaining analysis. Explicitly, in order to admit a fair comparison between a passive and active (aged) system, we should compare both systems at temperatures such that the relaxation times are of the same order of magnitude.
In the following, we will compare steady-state active and passive systems at different temperatures such that the relaxation times are of the same order of magnitude.
%typically use $T=0.3$ and $T=0.4$, or $T=0.25$ and $T=0.35$, for the active and passive case, respectively.

To quantify the degree of dynamic heterogeneity in passive and active glassy systems, we first benchmark the non-Gaussian parameter in the steady state \cite{kob1997dynamical,wisitsorasak2014dynamical}. Figure  \ref{fig:equilibrium}(b) shows the time dependence of $\alpha_2(t)$ at two  representative temperatures $T=0.3$ ($f=0.5$, $\tau_r=D_{r}=1$) and $T=0.4$ ($f=0$).
It can be seen that the $\alpha_2(t)$ of the active system starts to deviate from the passive case at a time $t \sim 1$. This time scale can readily be understood in terms of the ABP persistence time $\tau_r=D_r^{-1}=1$, i.e., the characteristic time scale at which the self propulsion force is expected to reorient and become prominently visible in the dynamics \cite{zottl2016emergent}.   
At intermediate times (t $>\tau_r$), both the passive and active system exhibit a peak in $\alpha_2(t)$, defining the time scale at which the Van Hove function has the strongest deviations from Gaussianity \cite{fernandez2020feedback}. In the passive system this peak is higher, meaning that the active system is less heterogeneous. 

To further study spatial correlations between the most mobile particles, one can compute the cluster size distribution \cite{kob1997dynamical,weeks2000three}. Here we identify mobile particles as those that move more than a distance $\Delta r=0.2$ in a time interval $\Delta t$. Two mobile particles belong to the same cluster if their distance is less than the average radius of the nearest neighbor shell $r_c$ (with $r_c$ determined from the radial distribution function). The cluster size distribution of such mobile particles thus gives direct insight into the real-space size of dynamically heterogeneous regions. The probability distribution $P(n)$ of clusters of size $n$ around the peak position is shown in the inset of Fig.\ \ref{fig:equilibrium}. In the passive case, the probability to find a bigger cluster is higher than in an active system. Therefore, we can conclude that the passive system is indeed more heterogeneous than the steady-state active system. Moreover, for both systems we have also verified that $P(n)$ is dominated by small clusters at short and long times, while bigger clusters can be found at intermediate time scales \cite{gebremichael2001spatially}.
\begin{figure}
 
  \includegraphics [width=9cm,height=3.3cm]{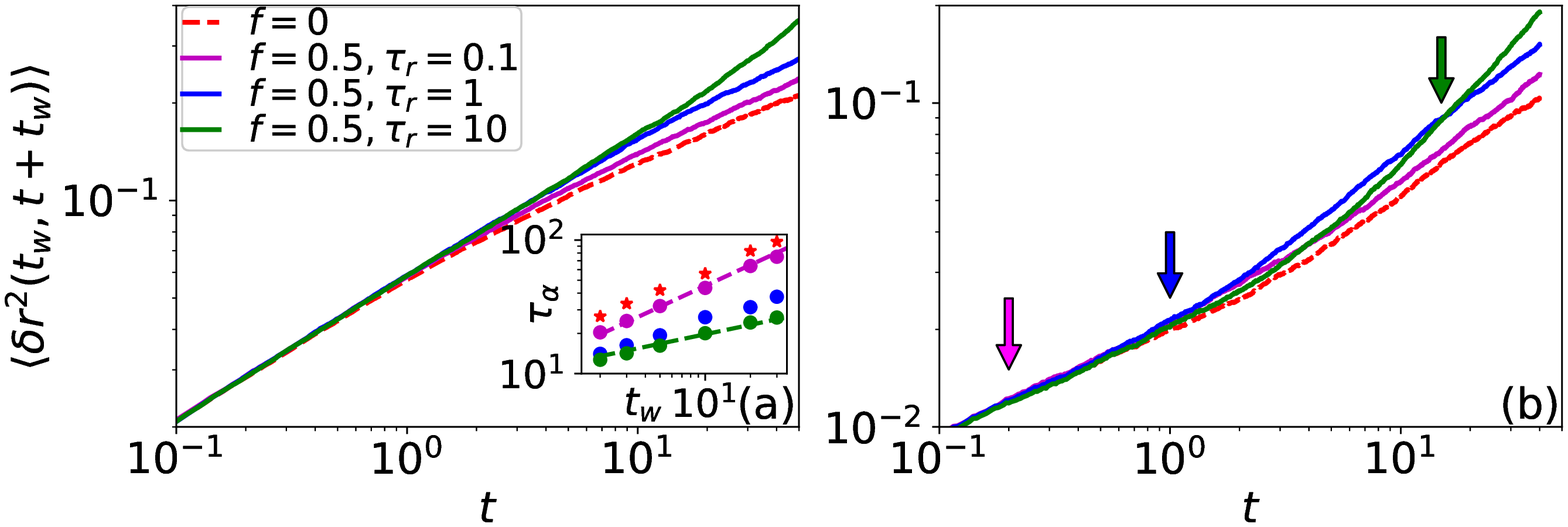}
  \caption{Mean-squared displacements as a function of time $t$ at $T_q=0.25$ for (a) $t_w=0$ and (b) $t_w=10$. The solid curves correspond to active systems with $f=0.5$ and different persistence times $\tau_r$; the dashed red line corresponds to a passive system. The inset of panel (a) shows the waiting-time dependence of the $\alpha$ relaxation time for different $\tau_r$; dashed lines represent a fit to the power law $\tau_{\alpha} \propto t_w^{\delta}$. For $\tau_r=0.1$ we obtain a fitted exponent of $\delta=0.52$, while for $\tau_r=10$ we find $\delta=0.24$.
   In panel (b) the arrows indicate which active system yields the highest MSD at the corresponding time $t$ (colors are the same as in panel a).}
  \label{fig:msd-dr}
\end{figure}

\textit{Aging is governed by a competition between thermal and active effects --}
To characterize the aging behavior, we first compare an active and passive thermal system quenched to the same temperature $T_q$. Figure \ref{fig:msd-dr} shows representative mean-squared displacements (MSDs) as a function of time $t$ for two waiting times $t_w$ and various persistence times $\tau_r$ at $T_q=0.25$. It can be seen that the MSD of the active thermal system is always equal to or higher than the MSD of the passive reference system; we have verified that this is true for all persistence times and waiting times studied (up to $t_w=1000$). To analyze the active aging dynamics in more detail, we distinguish between three regimes: a short-time regime $t \ll \tau_r$, a long-time regime $t \gg \tau_r$, and a non-trivial intermediate-time window  $t \sim \tau_r$ separating the short- and long-time dynamics. In the short-time regime, we find that the aged MSD of the active system is virtually indistinguishable from that of the passive reference system at the same $T_q$. This finding, which holds for all considered values of $\tau_r$ and $t_w$  and is also consistent with the short-time MSD of ideal ABPs in steady state \cite{bechinger2016active}, thus suggests that the short-time dynamics of an active thermal glass is fully governed by \textit{thermal} effects, regardless of the material's activity and age.

In the opposite limit of $t\gg \tau_r$, we observe a notable effect of the activity: The long-time dynamics is significantly sped up by the presence of active forces. This relative speedup depends monotonically on the persistence time, i.e.\ a larger $\tau_r$ (smaller $D_r$) yields faster dynamics. We also note that for the smallest persistence time considered here, $\tau_r=0.1$, the active system's MSD deviates only marginally from that of the passive reference system. This result holds regardless of the waiting time, and can be understood by the fact that a small $\tau_r$  causes particles to undergo fast reorientation, rendering the self-propulsion term less effective \cite{liluashvili2017mode,mandal2020multiple}. Overall, the above findings lead to our first main conclusion: The time-dependent aging dynamics of an active thermal glass is governed by a competition between thermal and active effects. Specifically, an active glass initially behaves very similar to a passive glass at the same temperature, but its long-time dynamics is controlled by activity.
Note that Mandal and Sollich \cite{mandal2020multiple} also reported an activity-dominated long-time regime for \textit{athermal} systems (referred to as 'ADA'); however, in contrast to Ref.\ \cite{mandal2020multiple}, the short-time aging dynamics in our case is inherently governed by the temperature.

To further quantify the long-time relaxation, we have also extracted the $\alpha$ relaxation times as a function of waiting time (inset of Fig.\ \ref{fig:msd-dr}(a)). This waiting-time dependence follows a power law $\tau_{\alpha} \propto t_w^{\delta}$ -- analogous to simple aging in passive systems \cite{kob1997aging} and in athermal active systems \cite{mandal2020multiple} -- with an exponent $\delta$ that decreases monotonically with $\tau_r$. Thus, as the system gets older, the active long-time dynamics will become increasingly faster compared to its passive analogue at the same $T_q$.

At intermediate times $t$, i.e.\ in the crossover regime between the temperature-dominated (short-time) and activity-dominated (long-time) regimes, the roles of activity and age become more complex. In particular we find that the ordering of MSD curves now depends on both $\tau_r$ and $t_w$. Careful inspection of Fig.\ \ref{fig:msd-dr} reveals that for $t_w=0$ (panel a) a larger $\tau_r$ always leads to a higher MSD, but for longer waiting times ($t_w=10$, panel b) this trend is violated. 
Specifically, for $t_w=10$, we find that when $t \sim 10^{-1}$ the active system with $\tau_r=0.1$ is the fastest (pink arrow in Fig.\ \ref{fig:msd-dr}(b)), but when $t \sim 1$ the system with $\tau_r=1$ becomes faster (blue arrow). Finally, when $t \gtrsim 10$ the system with $\tau_r=10$ yields the fastest dynamics (green arrow) and we recover the scenario of the long-time limit. We have verified that a similar  picture also applies for other finite waiting times (see supplementary material). We can rationalize this finding by noting that the persistence time $\tau_r$ is the intrinsic time scale of the active system, and the self-propulsion term can only start to become effective when the relevant time scale becomes comparable to $\tau_r$. In steady-state conditions this merely requires $t \sim \tau_r$, but in the case of aging the active system also needs to be sufficiently old ($t_w\gtrsim \tau_r$) to observe this effect.

\begin{figure}
    \centering
    \includegraphics [width=9cm,height=3.3cm] {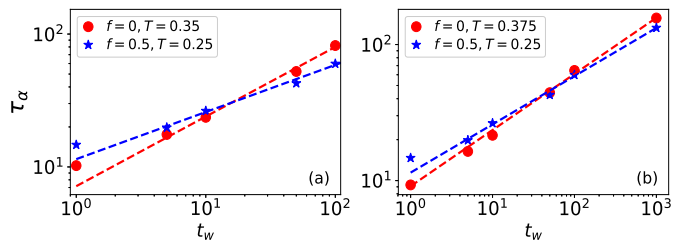} 
    \caption{Structural relaxation times $\tau_{\alpha}$ as a function of  waiting time $t_w$ for initial temperature $T_i=1$. The respective quenching temperatures for the passive ($f=0$) and active ($f=0.5, \tau_r=D_r=1$) system are (a) $T_q=0.375$ and $T_q=0.25$, and (b) $T_q=0.35$ and $T_q=0.25$. 
    The dashed lines represent a fit to the power law $\tau_{\alpha} \propto t_w^{\delta}$.}
    \label{fig:tau-Ti5}
\end{figure}

\textit{The 'effective temperature' changes with age --} 
We now seek to establish whether the long-time aging dynamics of an active glass can be mapped onto a passive system with a different effective temperature. To this end, we fix  $\tau_r=D_r=1$ and compare the active system to its passive counterpart at a higher quenching temperature $T_q$.
Figure \ref{fig:tau-Ti5} shows representative $\alpha$ relaxation times as a function of waiting time $t_w$ for two quenching temperatures $T_q$; the values of $T_q$ for the active and passive system are chosen such that the relaxation times in steady state are of the same order of magnitude. 
There are two important findings we infer from these data. First, the $\alpha$ relaxation times always follow a power law as a function of age, i.e.\ $\tau_{\alpha} \propto t_w^{\delta}$. The fitted exponents $\delta$ are also in good agreement with the values reported in Ref.\ \cite{mandal2020multiple} for athermal systems (see also supplemental material). 
Second, we find that for small $t_w$ the relaxation dynamics of the active system is generally \textit{slower} than the passive reference system, whereas for large $t_w$ the relaxation of thermal ABPs becomes \textit{faster}. Specifically, the active system at $T_q=0.25$ surpasses the passive curve for $T_q=0.35$ around $t_w\sim10$ (Fig.\ \ref{fig:tau-Ti5}(a)), while for the passive system with $T_q=0.375$ the crossing point lies near $t_w\sim100$ (Fig.\ \ref{fig:tau-Ti5}(b)). Hence, the simple picture of ABPs as hot colloids with a fixed effective temperature \cite{cugliandolo2019effective} breaks down during aging.

The above finding is in stark contrast with the athermal results of Ref.\ \cite{mandal2020multiple}, which showed virtually identical aging dynamics for active and passive glasses in the same parameter range ($\tau_r=1, T_q=0.37$). Our work thus implies that the addition of thermal noise in an active glassy system can change the aging dynamics profoundly, and leads to our second main conclusion: There is no rigorous mapping possible for the aging behavior of a thermal active system onto a passive system at a higher temperature, since the active system will always have a faster relaxation after a certain $t_w$. That is, the effective temperature of a thermal active glass evolves, and increases, with age.

To study the role of the initial temperature $T_i$, we have also tested the relation 
$F^{(T_{i,1})}(k,t_w,t+t_w) \sim F^{(T_{i,2})}(k,t_w+t_{(2,1)},t+t_w+t_{(2,1)})$, 
%$F^{(T_i=1)}(t_w,t+t_w) \sim F^{(T_i=5)}(t_w+t_{(5,1)},t+t_w+t_{(5,1)})$, 
for the intermediate scattering functions, with $t_{(2,1)}$ a fit parameter. This mapping has been previously reported for passive thermal systems \cite{kob1997aging}, and here we find that it also holds for our active thermal samples, at least for $T_{i,1}=1$ and $T_{i,2}=5$ (see supplementary material). 
We can rationalize this relation by considering that, as the system goes from $T_{i,2}$ to $T_q$, it will visit the configuration at $T_{i,2}>T_{i,1}>T_q$, so $t_{(2,1)}$ is the time needed for the system starting at $T_{i,2}$ to reach the same relaxation as the one found by a system equilibrated at $T_{i,1}$ at a fixed $t_w$ \cite{kob1997aging}. Overall, we can thus conclude that the effect of $T_i$ on the aging relaxation dynamics is qualitatively similar for thermal active and passive systems. % (see supplementary material).

%%%%%%%%%%%%%%%%%%%%%%%%%%%%%%

\begin{figure}
  \includegraphics[width=9cm,height=3.3cm]{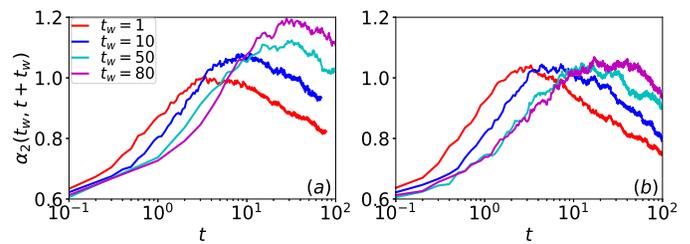}
  \caption{Non-Gaussian parameters as a function of time $t$ for waiting times $t_w=1,10,50,80$. (a) Passive system at quenching temperature $T_q=0.4$. (b) Active system ($f=0.5$, $\tau_r=D_r=1$) at $T_q=0.3.$}
  \label{fig:alpha2-0304}
\end{figure}

\textit{Non-Gaussianity and dynamic heterogeneity remain constant with age --}
We now turn to an aspect of glassy dynamics that has hitherto remained unexplored for aging active systems, namely dynamic heterogeneity.
Figure \ref{fig:alpha2-0304} shows how the non-Gaussian parameters $\alpha_2(t)$ for a passive and active thermal system, quenched to $T_q=0.4$ and $T_q=0.3$ respectively, evolve with $t_w$ as the system attempts to recover its steady-state behavior (cf.\ Fig.\ \ref{fig:equilibrium}(b)). In the passive case (Fig.\ \ref{fig:alpha2-0304}(a)), we find that for small $t_w$ the peak height of $\alpha_2(t_w,t+t_w)$ is  \textit{below} the steady-state value of $\alpha_2 \approx 1.2$, thus indicating that the passive relaxation dynamics is initially less heterogeneous than in steady state; as the system becomes older, the peak height grows and reaches the passive steady-state maximum after $t_w\approx80$. This gradual increase in heterogeneity is, however, strikingly different from the behavior we see in our active systems. Figure \ref{fig:alpha2-0304}(b) shows that in the active case the peak height of the non-Gaussian parameter remains remarkably constant with age, and is in fact always consistent with its steady-state value ($\alpha_2 \approx 1.1$). Thus, the degree of dynamic heterogeneity is manifestly age-independent for active thermal glass-formers. The only significant change with $t_w$ for our active system is that the peak position of $\alpha_2(t_w,t+t_w)$ shifts in time to ultimately recover the active steady-state dynamics. 
We have verified  for both the active and passive case that the same trend can be found for other values of $T_q$.

\begin{figure}
  \includegraphics[width=9cm,height=3.3cm]{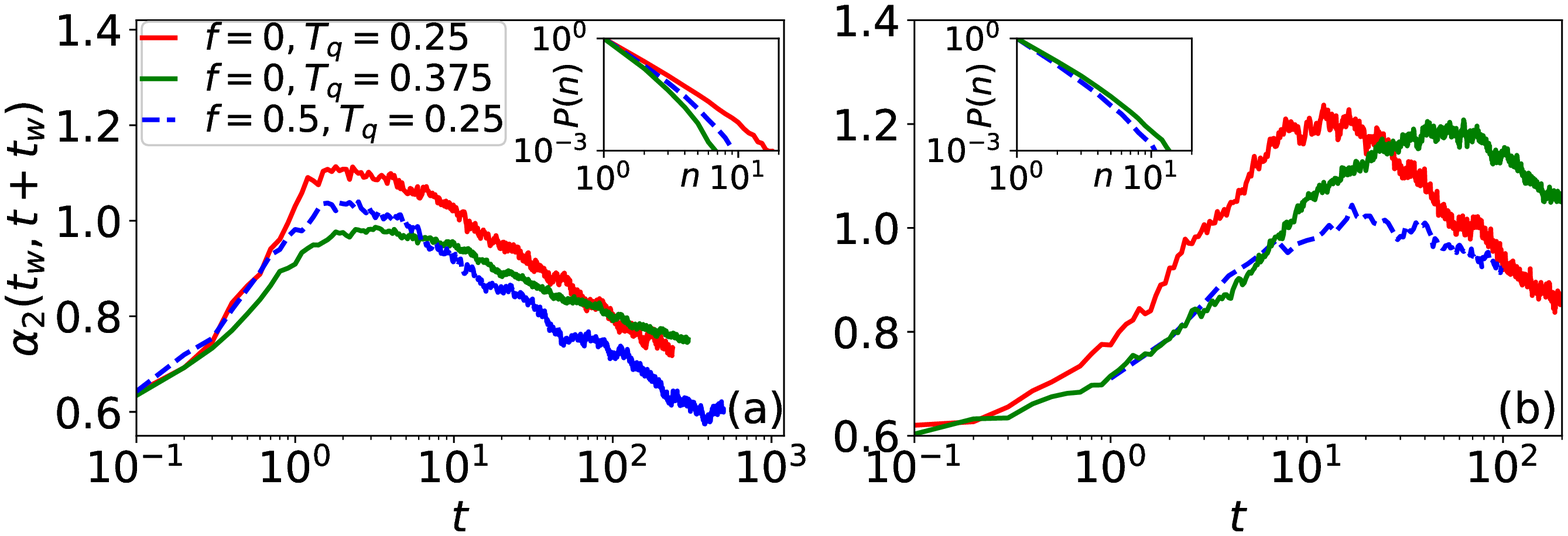}
  \caption{Non-Gaussian parameters as a function  of time $t$ for (a) $t_w=1$ and (b) $t_w=100$. The active system ($f=0.5$, $\tau_r=D_r=1$) is quenched to $T_q=0.25$, whereas the passive system is quenched to $T_q=0.25$ or $T_q=0.375$. The insets show the respective cluster size distributions $P(n)$ around the corresponding peaks of $\alpha_2(t_w,t+t_w)$.  
  }
  \label{fig:alpha2-037}
\end{figure}

Finally, let us examine the non-Gaussian parameters and cluster size distributions during aging for different temperatures. Figure \ref{fig:alpha2-037} compares $\alpha_2(t_w,t+t_w)$ for an active system at $T_q=0.25$ with a passive system at the same quenching temperature and at $T_q=0.375$. Here again we can identify a time-dependent competition between thermal and active effects: For sufficiently small $t_w$ (panel a) and small $t<\tau_r$ ($\tau_r=1$), the non-Gaussian parameters for the active and passive  samples at $T_q=0.25$ are very similar, but as time progresses the effect of activity becomes more prominent. To see this effect, let us first consider young glasses with $t_w=1$ (Fig.\ \ref{fig:alpha2-037}(a)). At intermediate time scales $t\gtrsim\tau_r$, the peak of $\alpha_2(t_w,t+t_w)$ is somewhat lower for the active system (Fig.\ \ref{fig:alpha2-037}(a)), implying that the active thermal sample undergoes less heterogeneous relaxation than its passive counterpart. However, compared to an equally young but higher-temperature passive system ($T_q=0.375$), the  active glass is slightly more dynamically heterogeneous, at least at the time scale where $\alpha_2(t)$ peaks. These differences are also reflected in the corresponding cluster size distributions (inset panel a). 

Importantly, as the waiting time increases ((Fig.\ \ref{fig:alpha2-037}(b)), we find that the active system at $T_q=0.25$ exhibits a significantly lower peak than both the $T_q=0.25$ and $T_q=0.375$ passive sample. This is in fact a direct consequence of the trend reported in Fig.\ \ref{fig:alpha2-0304}: In the active case the peak of the non-Gaussian parameter remains approximately constant with age, while in the passive case it grows. The measured size distributions of mobile clusters (inset panel b) also confirm that, for $t_w=100$, the largest correlated clusters are found for the passive system. Moreover, we have verified that in the passive case, regardless of $t_w$, the particle cluster size grows with decreasing temperature, consistent with the scenario in steady state \cite{donati1999spatial}. Overall, the above results unambiguously establish that, in terms of the microscopic relaxation dynamics, there is no simple mapping possible between an aging active thermal glass and a passive system with a different, but constant, effective temperature. We argue that the less pronounced, and approximately age-independent, degree of dynamic heterogeneity in active thermal glass-formers stems from the fact that the self-propulsion term enables particles to escape more easily, and more autonomously, from their local cages \cite{ni2013pushing}; this not only induces faster overall relaxation, but it also curtails the need for strongly cooperative, i.e., heterogeneous, relaxation dynamics.

\textit{Conclusions --} 
 In summary, our work reveals that thermal active glasses share some non-trivial aspects of aging phenomenology with athermal active and passive thermal glasses, but there are also several fundamental differences. Notably, the aging relaxation dynamics of active systems is governed by a time-dependent competition between thermal and active effects, with the ABP persistence time controlling both the time scale and magnitude of the activity-enhanced speedup in dynamics.  From our results it has not been possible to map the aging behavior of an active system onto a passive system at a different temperature; instead, we find that the manifested  effective temperature changes both with time and with age. Moreover, for an active thermal system, the dynamics is generally less heterogeneous than in the passive case, and the degree of non-Gaussianity remains remarkably constant with age. We hypothesize that both the relative speedup in dynamics and the relatively weak dynamic heterogeneity in aged active glass-formers is due to activity-enhanced cage breaking.  Finally, while we have focused on the aging dynamics following a temperature quench, we have verified that a similar phenomenology applies when using an activity-quench protocol (see supplementary material). In particular, at intermediate persistence times ($\tau_r\sim1$), quenching the temperature at fixed activity or quenching the activity at fixed temperature is essentially the same \cite{mandal2020multiple}.

\textit{Acknowledgments --} We thank Vincent Debets and Kees Storm  for their critical reading of the manuscript. This work has been financially supported by the Dutch Research Council (NWO) through a Physics Projectruimte grant.

 \bibliographystyle{apsrev4-1} 
\bibliography{./aging}
\end{document}